\documentclass[12pt]{iopart}
\usepackage{iopams}
\usepackage{graphicx,color,epsfig,rotate}

\newcommand{\egp}{e$_g^\prime$}

\begin{document}

\title{Nature of itineracy in CoV$_2$O$_4$: A first principles study}
\author{Ramandeep Kaur, Tulika Maitra$^{*}$  and  Tashi Nautiyal}
\address{Department of Physics, Indian Institute of Technology Roorkee, Roorkee- 247667,
Uttarakhand, India}
\ead{tulimfph@iitr.ac.in ($^{*}$corresponding author)}
\date{\today}


\begin{abstract}
Inspired by recent experiments, we have theoretically explored the nature of
itineracy in CoV$_2$O$_4$ under pressure and investigated, using first
principles density functional theory calculations, if it has any magnetic and
orbital ordering. Our calculations indicate that there could be two
possible routes to obtain the experimentally observed pressure induced metallicity in
this system. One is the spin-orbit interaction coupled with Coulomb correlation which can
take the system from a semiconducting state at ambient
pressure to a metallic state under high pressure. The other mechanism, as indicated by
our GGA+U calculations, is based on the presence
of two types of electrons in the system: localized and itinerant. An
effective Falicov-Kimball model could then possibly explain the observed insulator to metal
transition. Comparison of the
two scenarios with existing experimental observations leads us to believe that
the second scenario offers a better explanation for the mechanism
of insulator to metal transition in CoV$_2$O$_4$ under pressure.

\end{abstract}
\maketitle

\section{INTRODUCTION}

Orbital and magnetic order in spinel vanadates have recently become an intense
topic of discussion amongst the experimentalists and theoreticians alike
\cite{radaelli}. Several efforts have been made to unravel the complex
behavior involving lattice, spin and orbital degrees of freedom of such
systems\cite{motome}. In spinel vanadates, orbitally active vanadium ions
(V$^{3+}$) at the B-sites of AV$_2$O$_4$ form a frustrated
pyrochlore lattice. These systems often show multiple structural and magnetic
transitions at low temperatures\cite{reehuis,wheeler,garlea}. Things become
more interesting when the A-site is also occupied by one of the transition
metal (TM) ions which is magnetic. Competing exchange
interactions among the spins at $A$ and $B$ sites often result in more than
one magnetic transitions with change in temperature and at times in a complicated non-
collinear magnetic ordering at low temperatures \cite{garlea,chung}. Spinel
vanadates such as MnV$_2$O$_4$, FeV$_2$O$_4$ and CoV$_2$O$_4$ with magnetic
transition metal ions at A and B sites show a ferrimagnetic order of A and B
spins at low temperatures where all
the vanadium spins are parallel to one another but are antiparallel to those at
A sites\cite{tm-tsd,katsufuji}. Whereas on the other hand, the spinels such as ZnV$_2$O$_4$, CdV$_2$O$_4$
and MgV$_2$O$_4$ having non-magnetic A ion are found to have antiferromagnetically arranged
V spins \cite{reehuis,wheeler}. Usually the latter undergo structural
transition followed by a magnetic transition at low temperatures.

From literature it is clear that among the spinel vanadates, CdV$_2$O$_4$
resides on the
most insulating regime as the vanadium-vanadium distance d$_{V-V}$ in it is
the largest. Single crystal measurements \cite{kisma} reveal that
CoV$_2$O$_4$ has the shortest d$_{V-V}$ and hence is the closest to the
boundary between insulating and metallic regimes. Though this system is
semiconducting at ambient pressure, with the application of pressure which
reduces d$_{V-V}$ further, it crosses a critical value of d$_{V-V}$ and starts
conducting\cite{kisma}. However, the order of resistivities observed in the
pressure induced ``metallic'' regime is much higher\cite{kisma} in comparison to the
typical resistivities observed in case of a good metal. As far as the transport
property is
concerned, FeV$_2$O$_4$ lies in between CoV$_2$O$_4$ and MnV$_2$O$_4$.
In Table 1 we have summarized the available data from literature for these
AV$_2$O$_4$ vanadates \cite{reehuis,garlea,wheeler,tm-tsd,kisma,Giovannetti,Noriaki,kisma-thesis,rk-tm-tn,Pardo,huang,blanco}. We note that as we go from A= Mn to Fe to Co, the
d$_{V-V}$ decreases gradually and corresponding  magnetic ordering temperature
is found to increase. Also, as d$_{V-V}$ approaches the predicted
itinerant electron limit ($\sim$2.94 {\AA}) \cite{blanco}, the
tetragonal structural distortion decreases approaching the cubic structure.

CoV$_2$O$_4$ is a normal spinel having cubic structure with Fd$\bar{3}$m
symmetry where the V (3d$^2$) ions are placed within  O$_6$ octahedra and Co
(3d$^7$) ions are placed within O$_4$ tetrahedra\cite{huang}. Crystal field
produced by oxygen ions splits the d orbitals in $t_{2g}$ - $e_{g}$ manifolds.
The $t_{2g}$ ($e_{g}$) orbitals have lower energy than $e_{g}$ ($t_{2g}$)
orbitals in octahedral (tetrahedral) field. The O$_6$ octahedra are not the
ideal ones as the O-V-O angles deviate from 90$^o$ in this structure
\cite{kisma,huang}. This distorted octahedron causes a local trigonal
distortion which further splits the triply degenerate t$_{2g}$ orbitals of
vanadium into an a$_{1g}$ orbital and doubly degenerate e$_g^\prime$
orbitals. In accordance with Hund's rules, V$^{3+}$ in CoV$_2$O$_4$ has spin
S=1 with the electrons occupying  two out of the three t$_{2g}$ levels.

 \begin{table}
  \caption{Collected data for various vanadium spinels summarizing the trends. $T_S$ and $T_N$ define the structural and magnetic transition temperatures, respectively. Here AFM, FIM and FM stand for Antiferromagnetic, Ferrimagnetic and Ferromagnetic ordering, respectively. }
 \begin{center}
\begin{tabular}{|c|c|c|c|c|c|}
\hline
spinel system & d$_{V-V}$ ($\AA$) & T$_S (K)$ & T$_N (K) $  & c/a & source \\
\hline
CdV$_2$O$_4$ & 3.072 & 95 & 33 (AFM)   & 0.9877& ref. 9,10,11 \\
MnV$_2$O$_4$  & 3.014 & 56 & 56 (FIM)   & 0.98 & ref. 4,7,12 \\
&- & - & 53 (non collinear) & -  & ref 4,7 \\
FeV$_2$O$_4$ & 2.99 & 140 (HT)& 70 (FIM)  & 0.988 (HT)& ref. 12,13 \\
& &110 (LT) & - & 1.016 (LT) & 12,13\\
MgV$_2$O$_4$  & 2.976 & 65 & 42 (AFM)   & 0.9941 & ref. 5,14,15 \\
ZnV$_2$O$_4$ & 2.98 (in-plane) & 51 & 40 (AFM)  & 0.9949 & ref. 3,15\\
             & 2.97 (off-plane) &
              &    & & ref. 3,15 \\
CoV$_2$O$_4$ & 2.972 & -  & 152(FIM)  & 1 & ref. 9,16     \\
& - & - & 59 (non collinear) & -  & ref. 16 \\
\hline
\end{tabular}
\end{center}
\end{table}

Many vanadium spinels are observed to undergo two types of transitions as
temperature is lowered: first structural (e.g. cubic to tetragonal) and then
magnetic (e.g. paramagnetic to anti/ferrimagnetic) (see Table 1). The structural transitions
are often accompanied by an orbital order which then drives the magnetic order
at low temperatures. However, unlike other vanadium spinels, CoV$_2$O$_4$ is
reported to have no structural transition as the temperature is lowered
\cite{kisma}. Two magnetic transitions are observed \cite{huang} in
this system, one at
152 K from paramagnetic to collinear ferrimagnetic ordering between the Co and
V spins, and the other at 59 K from collinear ferrimagnetic to non-collinear
ferrimagnetic order. Hence
CoV$_2$O$_4$ makes an interesting study. In view of the experimental
observations
discussed above, many questions need to be explored
theoretically for this compound, like what
is the nature of itineracy? The effect of pressure on the electronic structure
of this compound, {\it primarily, the mechanism for insulator to metal transition under pressure},
nature of orbital order (if any), and the role of spin-orbit
interaction need to be investigated in detail. We have addressed these issues
in the present work using density functional theory (DFT) based calculations.
\begin{figure}
\begin{center}
\includegraphics[width=8cm]{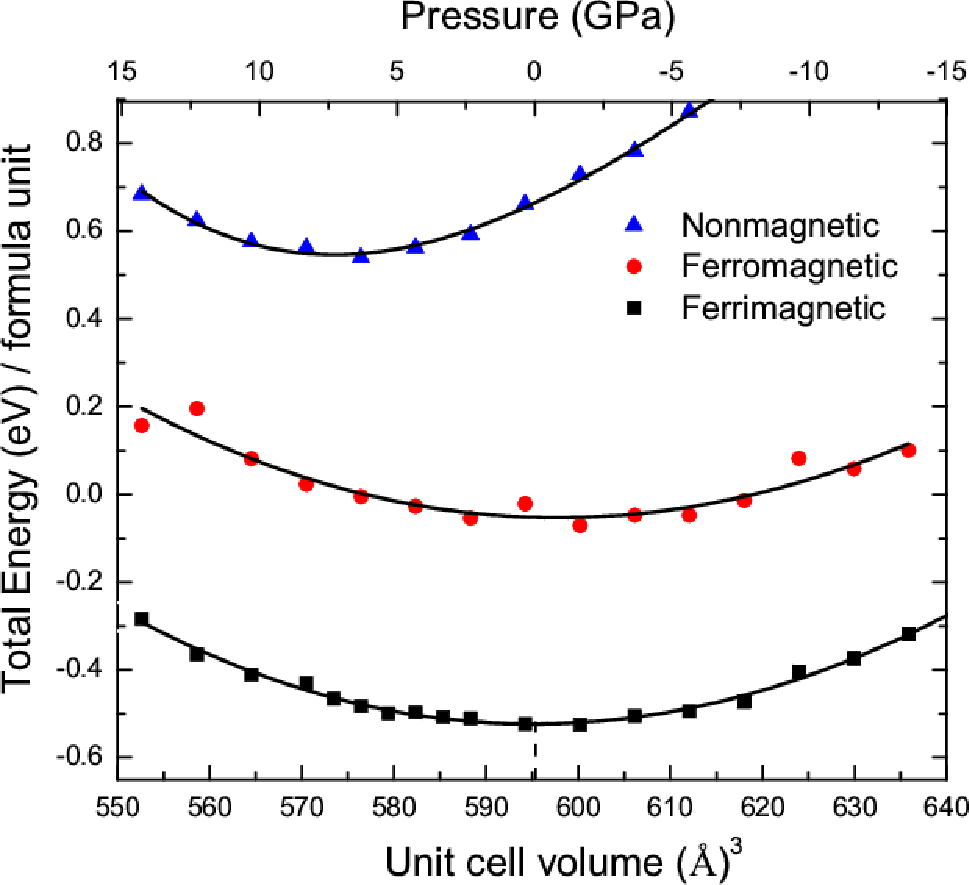}
\caption{Total Energy vs. conventional unit cell volume of
CoV$_2$O$_4$ within GGA. Base energy value -97779 eV is to be added to each. The corresponding pressure values are indicated on the top margin. Vertical line marks the equilibrium volume for the energetically favored FIM state.}
\end{center}
\end{figure}

\section{CALCULATIONAL DETAILS}
 As already stated, CoV$_2$O$_4$ has normal cubic spinel structure (with space
group Fd$\bar{3}$m (227)) at all the temperatures. It has two formula units
(f.u.) per primitive cell. Co ion is at (3/8,3/8,3/8) which is 8b
tetrahedral site, V ion sits at (0,0,0) which is 16c octahedral site, and O
ion sits at (x,x,x) which is 32e position. The structural
data for calculation are taken from experiment\cite{kisma}, where x=0.239.
The structure was then optimized using two different exchange correlation functionals: local spin density
approximation (LSDA) and Perdew-Burke-Ernzerhof (PBE) generalized gradient approximation (GGA)\cite{Perdew}, within full potential linearized
augmented plane wave method as implemented in WIEN2k code\cite{wien2k}. It was
observed that the optimized structure obtained within GGA is much closer to
the experimental structure than that obtained within LSDA. Hence further
calculations were performed using the
GGA for exchange correlation. The
muffin-tin radii were set to 1.91, 1.94 and 1.69 {\AA} for Co, V and O,
respectively. The plane-wave cut off (R$_{mt}$.K$_{max}$) was set to 8.0,
and 72 $\vec{k}$ points were used in the irreducible wedge of the Brillouin
zone. The expansion of the radial wave function in spherical harmonics
 was taken up to angular momentum quantum number $l$ = 10.
We also performed more refined calculations by including the correlation
effects arising from the d electrons of Co and V. For this, we employed
self-interaction corrected (SIC) GGA+U approach\cite{anisimov} which takes
into account the onsite Coulomb interaction U and removes the self-Coulomb
and self-exchange correlation energy. Spin orbit (SO) coupling is included by
the second-variational method\cite{koelling} with scalar-relativistic wave
functions. In this approach, firstly the wave functions are calculated for
both the spin directions without involving SO interaction term in the eigen
value problem. Then those functions are used as basis functions to solve the
new eigenvalue problem which has the SO interaction term in the total
hamiltonian.

\section{RESULTS AND DISCUSSION}

\subsection{Magnetic ordering}
\begin{figure}
\begin{center}
\includegraphics[width=8cm]{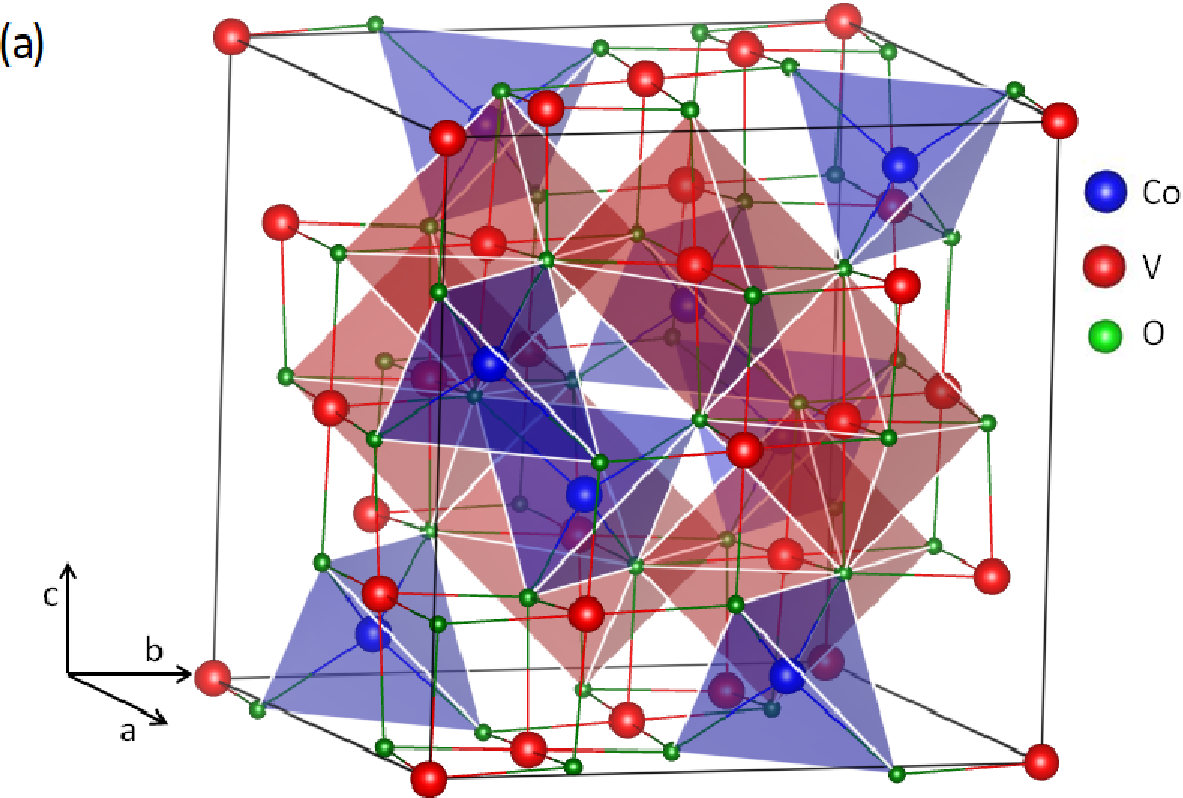}
\includegraphics[width=3cm]{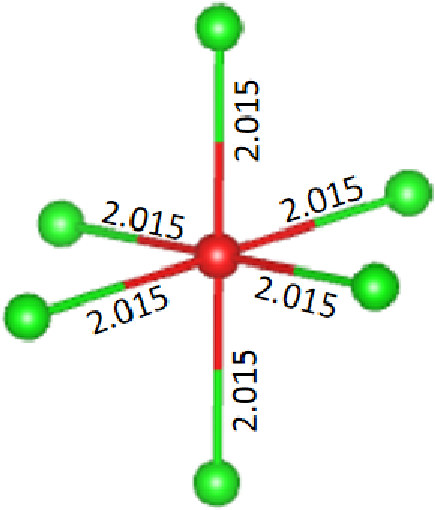}
\caption{(a) The cubic spinel unit cell for CoV$_2$O$_4$. Red (Blue)
oxygen octahedra (tetrahedra) are around V (Co) ions. (b) The VO$_6$ octahedron with bond lengths in {\AA}}
\end{center}
\end{figure}

Pursuit of a better understanding of the recent experimental observations led us to move to higher levels of sophistication of the calculations: from GGA to GGA+U to GGA+U+SO approximations.
In the following, we present the electronic structure calculations within the above approximations and the analysis of the
results to propose possible underlying mechanisms to explain the recent experimental findings \cite{kisma} regarding the effect of pressure on the transport
properties of this system. In the process we also studied the effect of pressure on the magnetic and orbital order
(if any) of the system. Firstly DFT calculations using
GGA were performed to ascertain the magnetic ground state and to obtain the
optimized
lattice parameters, characterized by the minimum total energy, of CoV$_2$O$_4$. Fig. 1 shows the total energy per formula unit as a function of unit cell volume (and of pressure)
for Nonmagnetic (NM) or spin unpolarized, Ferrimagnetic (FIM) and Ferromagnetic (FM)
arrangements
of Co and V spins. We note that for the FIM ordering, Co(V) spins within
A(B) sublattice are arranged ferromagnetically, and the two sublattices are
ordered antiferromagnetically with respect to each other. It is
clearly seen that this material has a strong preference for
magnetic states over the NM state since the total energy for the NM state is
the highest of all in the studied volume range. It also shows that
ferrimagnetic arrangement of Co and V spins constitutes the ground state as the energy of ferromagnetic solution is found to be higher by 0.502 eV/f.u. at
the equilibrium volume (= 595.440 {\AA}$^3$). Our calculations predict the equilibrium
lattice constant to be 8.4129 {\AA}, larger only by 0.2\% compared to the
experimental value ({8.4073 \AA})\cite{kisma}.
We have also calculated the total
energy for antiferromagnetic arrangement of vanadium spins as observed in some
other spinel vanadates like ZnV$_2$O$_4$ and MgV$_2$O$_4$\cite{reehuis,wheeler}. We find that the total energy for such a state at the equilibrium
volume is higher than that for the ferrimagnetic case by 0.181 eV/f.u. Therefore our calculations give FIM ordering as the most favoured one, in agreement with experiments. We have also calculated the equation of state and obtained the bulk modulus and its pressure derivative for the FIM ordering which are found to be 202.23 GPa and 4.09, respectively. These values are comparable to those for other spinel systems \cite{Errandonea}. 

Cation inversion is an interesting phenomenon that is exhibited by some Co spinels such as CoFe$_2$O$_4$ and CoAl$_2$O$_4$ \cite{CoFe2O4,CoAl2O4}. To explore this possibility, we studied the cation inversion in CoV$_2$O$_4$ and found that the inverse spinel structure with Co and V at octahedral sites and V at tetrahedral sites has higher energy than the normal spinel structure by 1.126 eV/f.u. This implies that cation inversion is energetically not favoured in CoV$_2$O$_4$ as Co and V ions prefer tetrahedral and octahedral sites, respectively. Hence in the following we present our results for the normal spinel structure with optimized unit cell and in the ferrimagnetic ordering.
\subsection{GGA calculations}
Fig. 2 shows the unit cell for CoV$_2$O$_4$ with VO$_6$ octahedra and CO$_4$ tetrahedra shaded in different colors. All the Co-O distances and O-Co-O
angles have same value equal to 1.981 {\AA} and 109.47$^o$, respectively. However, the
octahedron around V ions is not perfect due to the presence of trigonal
distortion. The O-V-O angles deviate by approximately 5$^o$ from the perfect
90$^o$, having two different values 84.62$^o$ and 95.38$^o$, though all the
V-O distances have same value, equal to 2.015 {\AA} (Fig. 2 (b)). The d$_{V-V}$
distance for the optimized unit cell is 2.974 {\AA} which is fairly close to
the limit of itineracy for vanadium spinels predicted from previous studies\cite{blanco}.  In Fig. 3 we present the partial density of states (DOS) for various d states of vanadium within GGA, as well as the total electronic DOS for CoV$_2$O$_4$ and of Co, V and O (inset). It
clearly shows that the GGA calculations predict a metallic ground state for
CoV$_2$O$_4$ with a nonzero DOS at the Fermi level (E$_F$). This is not in agreement
with the experimental findings which designate this material as a semiconductor
at ambient pressure\cite{kisma}. It is worth emphasizing that in case of the vanadium ions, a local trigonal distortion is also present, in addition to the octahedral crystal field due to
surrounding oxygens. This trigonal
distortion  further splits the lower energy
triply degenerate t$_{2g}$ orbitals into an a$_{1g}$ (d$_{z^2}$) orbital and
the doubly degenerate e$_{g}'$  orbitals, with the latter having higher energy. We observe from GGA DOS that
both a$_{1g}$ (d$_{z^2}$) and e$_g^\prime$ states are
present at the E$_F$. We note that both Co and V are
transition metal ions (having partially filled d orbitals)
and are, therefore, expected to be strongly correlated. Hence in order to achieve the experimentally observed
semiconducting state, we included the
Coulomb correlation in our calculations within GGA+U approximation\cite{anisimov}.


\begin{figure}
\begin{center}
\includegraphics[width=8cm]{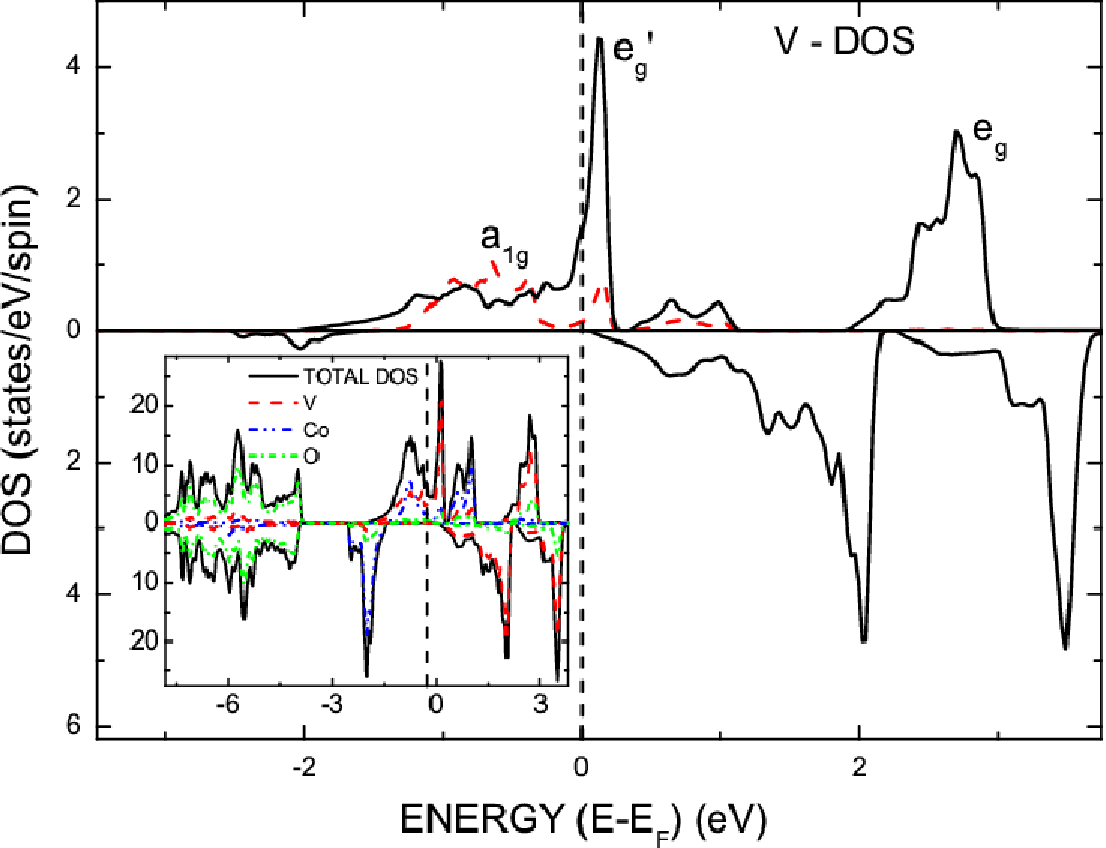}
\caption{ Spin polarized DOS of vanadium within GGA; inset shows the total DOS for CoV$_2$O$_4$ and of Co, V and O. }
\end{center}
\end{figure}

\subsection{GGA+U: Magnetic, orbital and transport properties}

In this section we discuss our results obtained within GGA+U approximation
and compare the same with the experimental observations.
In an effort to see the impact of correlation
on the results, we tried various values (ranging from 2 to 5 eV) of U$_{eff}$ (=U-J,
where U is the Hubbard parameter representing the on-site Coulomb interaction,
and J is the Hund's exchange interaction strength) for both Co and V. Here, we present the
results with U$_{eff}$ equal to 4 eV for Co, this also being the usual choice of U$_{eff}$ for Co systems\cite{khomskii}. As discussed in detail below, with
this value of U$_{eff}$ for Co, we observe that the occupied Co d states are
pushed down far from the E$_F$ leaving mainly the V d states at
and around E$_F$. In Fig. 4 we show the effect of $U$ on the vanadium d states
in a range of U$_{eff}$ values which are found relevant also for other vanadium
spinels\cite{tm-tsd,tm-rv}. We discuss later in this section some important
physical insights that one can get from Fig. 4 about the possible microscopic
mechanism
behind the transport properties of this system.
We note that the ferrimagnetic
arrangement of Co and V spins is found to be the ground state within
GGA+U also, with the total energy of this phase being lower than
that of the corresponding FM phase by 0.145 eV/f.u.

\begin{figure}
\begin{center}
\includegraphics[width=8cm]{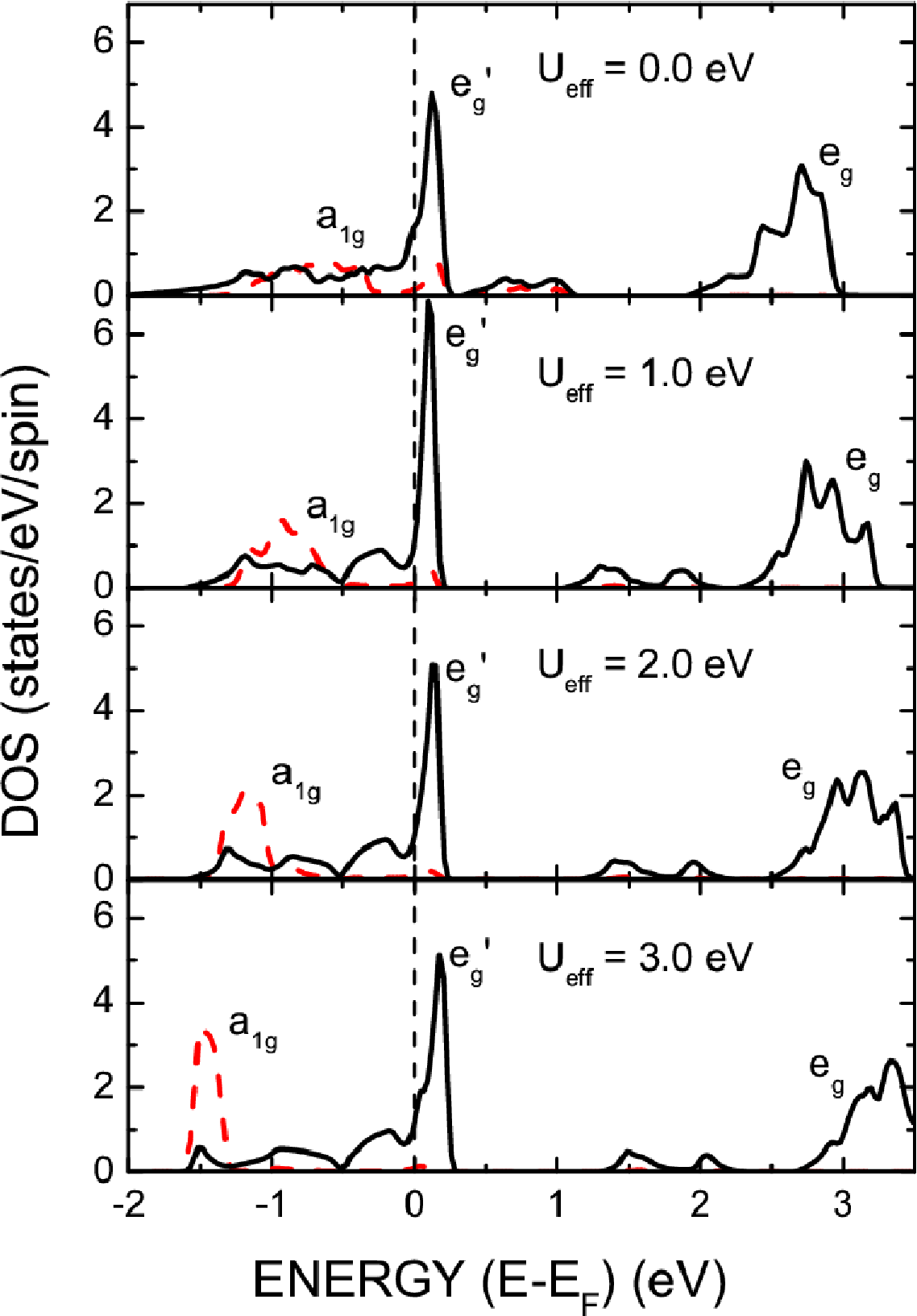}
\caption{ Spin up DOS for vanadium d states around Fermi level within GGA and GGA+U (with U$_{eff}$=1, 2 and 3 eV) showing increasing localization of a$_{1g}$
states with U.}
\end{center}
\end{figure}

Fig. 5(a) shows the electronic DOS within GGA+U. The top panel displays the
total
DOS for Co, V and O; middle and the bottom panels show the DOS for the d states
of Co and V, respectively. Tetrahedral crystal field lifts the degeneracy of
Co d-orbitals into e$_g$ (i.e. d$_{z^2}$ and d$_{x^2-y^2}$) states with lower
energy, and t$_{2g}$ (i.e. d$_{xy}$, d$_{yz}$, d$_{zx}$)  states with higher
energy. In the majority (down) spin sector, both e$_g$ and t$_{2g}$ orbitals
are filled with 5 electrons and the remaining two electrons of Co$^{2+}$ fully occupy the
lower lying doubly degenerate e$_g$ orbitals in the minority (up) spin channel.
So there are no Co states present at E$_F$, and with the application of U,
the fully occupied d states of Co are further pushed down with respect to E$_F$.

Comparing our results for V ion with the
GGA calculations (discussed in the previous section), we observe that with the
application of U, the a$_{1g}$ band is pushed below E$_F$ and with increase in U, goes farther down becoming more and more narrow (localized). On the other hand, the doubly degenerate e$_{g}'$ bands
remain delocalized and pinned at the E$_F$ (see Fig. 4). The spin-split a$_{1g}$ band becomes
fully occupied by one electron of vanadium. In this
case (Fig. 5(a))
with U$_{eff}$ = 2.7 eV\cite{footnote}, the orbital lies well below E$_F$ (by about 1.5 eV).
The e$_{g}'$ bands, which are now seen straddling the E$_F$, are filled only
partially by the 2$^{nd}$ vanadium electron.
The system is hence found to be metallic even after taking into account reasonably
 strong Coulomb correlation in the calculations. However, an important
observation one can clearly make from these results is
that the system contains two types of electrons: localized and
itinerant, as also predicted from a recent experiment\cite{kisma} on CoV$_2$O$_4$ and in accordance with previous
theoretical results on similar spinel vanadates\cite{blanco}. The GGA+U
DOS for a$_{1g}$ and e$_{g}'$ orbitals shown in Fig. 4 strongly suggests that this system can be modelled by a spinless Falicov-Kimball model (FKM) with two degenerate itinerant bands with the Hamiltonian given below

\begin{eqnarray}
{H} =-\sum_{\langle ij\rangle,\alpha=1,2}(t_{ij}^\alpha+\mu\delta_{ij})d^{\dagger}_{i\alpha}d_{j\alpha}
+E_f \sum_{i} (f^{\dagger}_{i}f_{i})
+U \sum_{i,\alpha=1,2}{(f^{\dagger}_{i}f_{i}d^{\dagger}_{i\alpha}d_{i\alpha})}
\end{eqnarray}
where the a$_{1g}$ state is
localized and fully occupied by one electron and the doubly degenerate
e$_{g}'$ states are itinerant and half filled by one electron\cite{umesh2}.
Here $d^{\dagger}_{i\alpha}, d_{i\alpha}$  are, respectively, the creation and
annihilation operators for itinerant e$_{g}'$ electrons and $f^{\dagger}_{i},
f_{i}$ are the same for localized a$_{1g}$ electrons. $\alpha=1$ and $2$
represent the two degenerate e$_{g}'$ orbitals. The first term in Eq.(1) is the kinetic energy of e$_{g}'$ electrons. The second term represents the dispersionless energy level $E_{f}$ of the a$_{1g}$ electrons while the third term is the on-site Coulomb repulsion between $d$ (or \egp) and $f$ (or a$_{1g}$) electrons.

As is well known\cite{falicov}, FKM can give rise
to an insulating state (as observed under ambient pressure in CoV$_2$O$_4$) due
to the strong scattering of the itinerant electrons from the localized
ones. The same physics will work even in the presence of two degenerate
itinerant bands as long as there is no quantum mixed valence. For infinite
dimensions, solutions on FKM~\cite{kotliar} reveal that the local
inter-orbital susceptibilities, $\chi_{\alpha,f}(\omega)$, diverge with a power
law, underlining an incoherent metallic (`bad metal') state. On the application of
pressure, the broadening of both itinerant and localized bands would give rise to a small
hybridization
between the two i.e. the excitonic amplitude such as $<f^\dagger_id_{i\alpha}>$
then becomes finite (i.e. quantum mixed valence). In that case, it is known that
the above model, Equation (1), containing such a hybridizing term can lead
to a correlated metal with bad metallic properties\cite{umesh2,AT-TiSe2}. This could be
a possible scenario behind the metallic behavior (with high resistivities) under
application of pressure.
More sophisticated calculations (beyond DFT) such as Dynamical Mean Field Theory (DMFT)
\cite{dmft} could provide useful insight on this mechanism.
\begin{figure}
\begin{center}
\includegraphics[width=14cm]{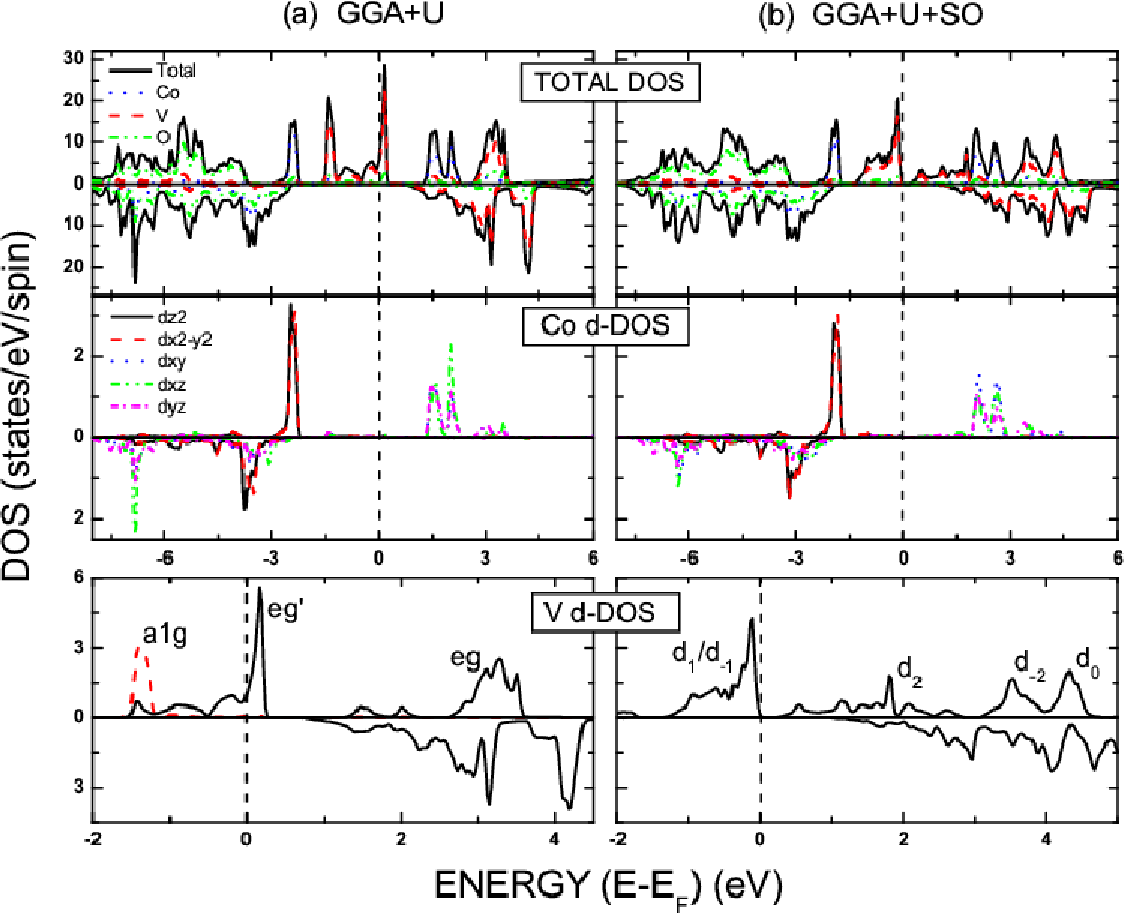}
\caption{ Spin polarized DOS within (a) GGA+U and (b) GGA+U+SO around the Fermi level (U$_{eff}$ = 2.7 eV).}
\end{center}
\end{figure}

\subsection{GGA+U+SO: Transport properties under pressure}

In an effort to obtain the experimentally observed semiconducting nature of
CoV$_2$O$_4$ within the limits of DFT, we included the SO coupling in our GGA+U calculations. The SO
coupling is known to lift many degeneracies and thus is likely to give rise to
the desired gap in the energy spectrum of conduction electrons. Also, it is
often found to be significant and has an appreciable effect on the orbital
order in other vanadium spinels\cite{rk-tm-tn,tm-rv}. In Fig. 5(b) we show
the DOS calculated within GGA+U+SO approximation with the same U$_{eff}$ (= 2.7 eV). Magnetization direction
is taken to be along (001). What one observes from this
DOS is that the complex t$_{2g}$ orbitals are now split into $d_1/d_{-1}$
and $d_{2}$. The former has lower energy, lies below the
E$_F$ and is occupied with the two vanadium electrons, whereas the latter
having higher energy lies above E$_F$ and remains unoccupied. The real d-orbitals are defined in terms of these complex orbitals as d$_{xy}$ = d$_{-2}$-d$_2$, d$_{xz}$ = d$_{-1}$-d$_1$, d$_{yz}$ = d$_1$+d$_{-1}$, d$_{x^2-y^2}$ = d$_2$+d$_{-2}$ and d$_{z^2}$ = d$_0$.
It is encouraging to note that this splitting in presence of SO coupling gives the desired gap at the E$_F$.
This energy gap created by SO coupling is about 0.145 eV at ambient
pressure which is typical of a semiconductor.
More importantly, our results indicate that the inclusion of SO
coupling along with the Coulomb correlation is essential to obtain correct
ground state of CoV$_2$O$_4$ even
though it contains only 3-d transition metal elements Co and V. In view of DFT
calculations  underestimating the band gap, the actual band gap of
CoV$_2$O$_4$ could be greater than 0.145 eV. There is no experimental
data available. However, knowing that CoV$_2$O$_4$ has d$_{V-V}$ close to the
limit of itineracy for these spinel vanadates, we do not expect this band gap
to be very large either. It is interesting to note that the $d_{2}$
band which lies just above the E$_F$ is very wide (bandwidth about 2.5 eV) compared to the occupied $d_1/d_{-1}$ bands just below  E$_F$  (bandwidth about 1 eV).
\begin{figure}
\begin{center}
\includegraphics[width=9cm]{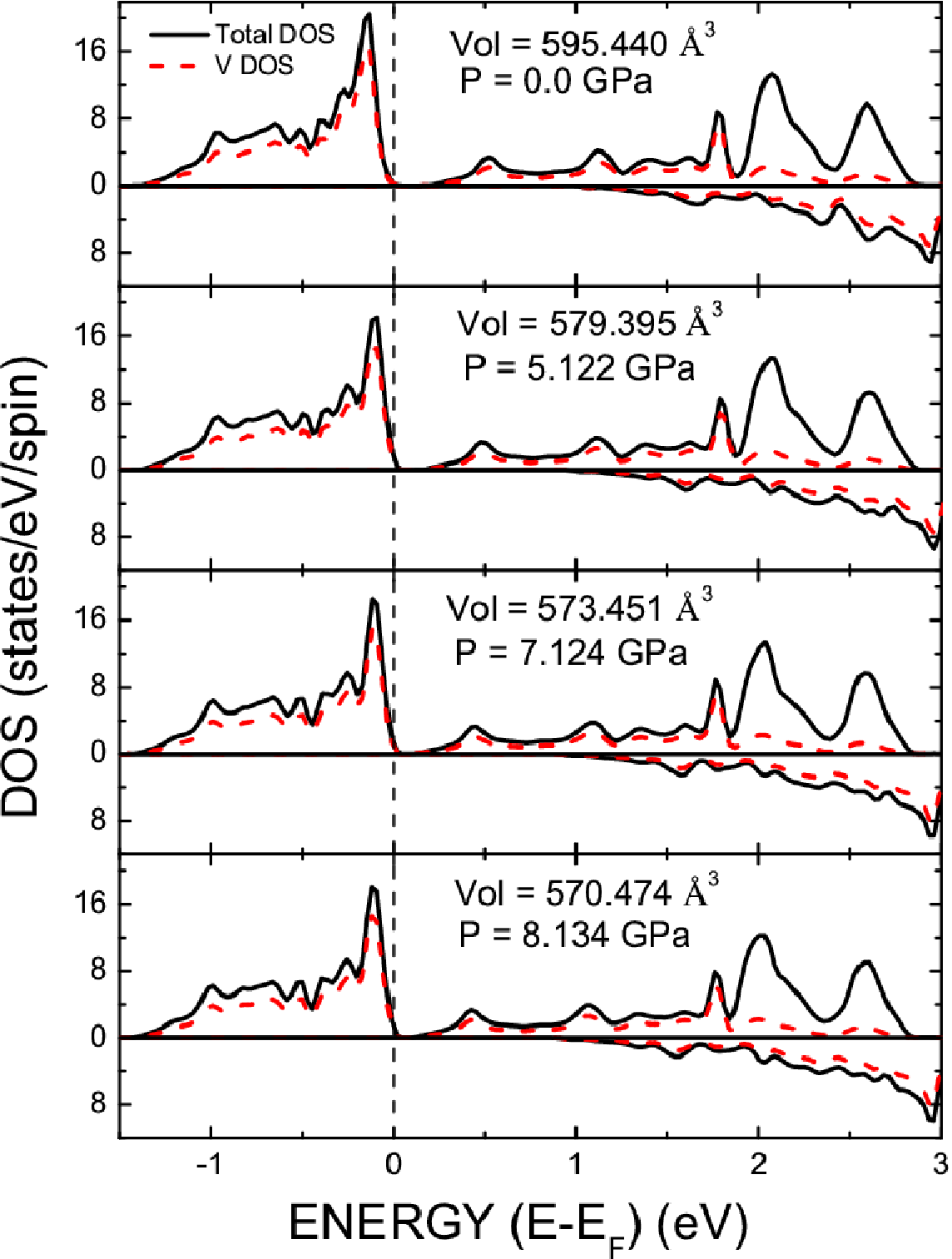}
\caption{ Total DOS and V DOS as a function of energy within GGA+U+SO (U$_{eff}$=2.7 eV). The pressure (P) and corresponding volume (Vol) values are listed for each panel. The vertical line marks the E$_F$.}
\end{center}
\end{figure}

In recent experiments on CoV$_2$O$_4$, Kismarahardja et al.
\cite{kisma} have studied the impact of pressure on the transport
properties of this system. They observed that on applying pressure, this
system undergoes a semiconductor to metal transition. Though at low temperatures
the system remains semiconducting under higher pressures, the metallic
character is seen to appear in some intermediate temperature range below T$_C$ where the system has collinear ferrimagnetic order.
However, the resistivities observed in the ``metallic" phase are an
order of magnitude higher than that of typical metals. In fact, the resistivities
of the metallic and insulating states are almost comparable in magnitude, only
the slopes of resistivity versus temperature have different signs in these two
states. The situation is somewhat akin to many of the `bad metals' found in the
literature where the resistivity in the putative metallic state is nearly the
same as in the insulating state. This could be a signature
of correlated metals (as in other `bad metals') where strong scattering
due to electronic correlations dominates the transport.
To check if we can recover this transition from our theory, we
carried out calculations within GGA+U+SO at different pressures. 
At this point, we would like to mention that certain cubic spinels such as ZnGa$_2$O$_4$ \cite{Errandonea} are observed to undergo cubic to tetragonal 
structural transition under pressure. Since no experimental
information is available in literature about the                          
distortions introduced by application of pressure on cubic spinel CoV$_2$O$_4$, it is worth
exploring the possibility of a cubic to tetragonal transition in this system on
increasing the pressure. In this context, we performed total energy calculations (within GGA) 
for the tetragonal structure at 8.134 GPa pressure (close to the highest pressure 
studied in experiments by Kismarahardja et al.\cite{kisma}) with both 
c/a $<$ 1 and c/a $>$ 1. We observed that both these tetragonal structures have total energy greater than that for the cubic structure with c/a=1. This eliminates the possibility of a cubic to tetragonal 
structural
transition under pressure in CoV$_2$O$_4$ in the pressure range studied below. We also note that cubic to tetragonal transition observed by D. Errandonea et
al. in ZnGa$_2$O$_4$ occurs at 31.2 GPa which is much higher than the pressure 
range studied in this work. Our interest lies in explaining the experimentally observed change in resistivity in CoV$_2$O$_4$ upto 8 GPa, hence we have not explored the higher range of pressure.

The unit cell volumes, all the relevant bond lengths and the corresponding band gaps at various pressures are listed in Table 2. All the bond lengths are seen to decrease with pressure. As the application of pressure in this case does not change the symmetry, the fractional coordinates (x,x,x) of oxygen remain same with increasing pressure. The effect
of pressure on the total DOS of CoV$_2$O$_4$ is presented in Fig. 6. We can
clearly see that with the application of pressure, band gap decreases. It would be interesting to compare this pressure dependence with that of orthovanadates (AVO$_4$) where such a study was recently reported\cite{Panchal}. Unlike the orthovanadates, we do not see any structural phase transition in the range of pressure studied here. Therefore no discontinuity is observed in band gap versus pressure plot unlike the case of orthovanadates. The calculated pressure coefficient (dE$_g$/dP) for CoV$_2$O$_4$ is found to be about -14.5 meV/GPa. This is comparable to the values measured experimentally in scheelite structures of AVO$_4$ \cite{Panchal}. 

As the band gap decreases significantly at higher pressures (for example,
at $\simeq$ 8 GPa, the band gap is $\sim$ 27 meV), the
system can achieve activated conduction due to thermal
excitations across the band gap. As mentioned earlier, the conduction
band just above the E$_F$ is highly dispersive compared to the valence band.
From the band structure (not shown here) also, we see this feature.
So the ``metallic" behaviour (with relatively large resistivity) observed in the experiments\cite{kisma} could be due to the
thermal excitations of electrons from the narrow valence band
just below the E$_F$ to highly dispersive conduction band just above
the E$_F$. This would, although, essentially be an activated conduction unlike a true metal.
The slope of resistivity versus temperature graph will, thus, not change sign unlike that observed experimentally.
With further increase in pressure, however, the band
gap is expected to close due to the overlap of valence and conduction bands.
The system in this case will be highly metallic (even at low temperatures) because of the highly
dispersive nature of the conduction band. This is again not consistent with the experimental observations
of insulating state at low temperatures and metallic state with high restitivies at intermediate temperatures. So the results presented in this and the
previous subsections strongly indicate that the insulator to metal transition
observed experimentally in this system under pressure could be better explained through a mechanism involving both itinerant and localized electrons described by Equation (1).

Regarding possible orbital order in this system, we would like to mention that we do not see any orbital ordering as both $d_1/d_{-1}$ orbitals, below the E$_F$, are
degenerate and equally occupied by the two vanadium electrons. Net orbital moment at each
vanadium site is also observed to be nearly zero.

 \begin{table}
  \caption{Variation of unit cell volume, bond lengths (d$_{V-V}$, d$_{Co-O}$, d$_{V-O}$) and band gap  with change in pressure using GGA+U+SO }
 \begin{center}
\begin{tabular}{|c|c|c|c|c|c|}

\hline
Pressure (GPa) & Volume ({\AA}$^3$) &d$_{V-V}$ ({\AA}) &d$_{Co-O}$ ({\AA}) &d$_{V-O}$ ({\AA}) & Band Gap (eV) \\
\hline
0.00 & 595.440 & 2.974 & 1.981 & 2.015 & 0.145  \\
5.122 & 579.395 & 2.947 & 1.964 & 1.997 & 0.085 \\
7.124 & 573.451 & 2.942 & 1.960 & 1.994 & 0.053 \\
8.134 & 570.474 & 2.932 & 1.954 & 1.986 & 0.0271 \\
\hline

\end{tabular}
\end{center}
\end{table}


\section{CONCLUSIONS}
Our DFT calculations show that the consideration of Coulomb correlation alone is not
enough to obtain the semiconducting nature of CoV$_2$O$_4$ at ambient
pressure. One has to consider SO interaction in order to reproduce
the experimentally observed semiconducting behavior. Our GGA+U+SO results
indicate the presence of a highly dispersive conduction band just
above the E$_F$ separated from a narrow valence band by a small band gap.
With the application of
pressure on the system, the band gap is observed to reduce significantly
bringing down the wide conduction band very close to the valence band maximum.
Therefore, at higher pressures the thermal excitations from the narrow valence
band to wide conduction band can lead to an activated conduction. However,
the experimental observations that the resistivity of metallic and insulating
phases being of the same order clearly indicates that this is a correlated metal
rather than simple metal with free electrons. This picture is also borne out
in our GGA+U calculations where we do see the presence of two types of
electrons: itinerant and localized. So we believe that the insulator to metal
transition in this system under pressure can be better explained from an effective
spinless Falicov-Kimball model with two degenerate itinerant bands and one localized band as discussed.

 \section{ACKNOWLEDGEMENT}

\noindent This work is supported by CSIR, India project grant
(ref. no. 03(1212)/12/EMR-II). RK acknowledges CSIR (India) for a research fellowship.


\begin{thebibliography}{99}
\bibitem{footnote} U$_{eff}$=2.7 eV was chosen in order to ensure that the system
remains insulating (consistent with experimental observations) at zero temperature within GGA+U+SO at the highest pressure($\simeq$ 8 GPa) considered in the experiments\cite{kisma}.
\end{thebibliography}
\end{document}